# Electric-field fluctuations as the cause of spectral instabilities in colloidal quantum dots

Frieder Conradt[1], Vincent Bezold[1], Volker Wiechert[1], Steffen Huber[3], Stefan Mecking[3], Alfred Leitenstorfer[1], and Ron Tenne[1]

[1] Department of Physics and Center for Applied Photonics, University of Konstanz, D-78457 Konstanz, Germany

[2] Chair of Chemical Materials Science, Department of Chemistry, University of Konstanz, D-78457 Konstanz, Germany

## Abstract
Spectral diffusion (SD) represents a substantial obstacle towards implementation of solid-state quantum emitters as a source of indistinguishable photons. By performing high-resolution emission spectroscopy for individual colloidal quantum dots at cryogenic temperatures, we prove the causal link between the quantum-confined Stark effect and SD. Statistically analyzing the wavelength of emitted photons, we show that increasing the sensitivity of the transition energy to an applied electric field results in amplified spectral fluctuations. This relation is quantitatively fit to a straightforward model, indicating the presence of a stochastic electric field on a microscopic scale whose standard deviation is 9 kV/cm, on average. Compensating the commonly observed intrinsic electric bias with an external one, we find that SD can be suppressed by up to a factor of three in CdSe/CdS core/shell nanorods. The current method will enable the study of SD in multiple types of quantum emitters, such as solid-state defects or organic lead-halide perovskite quantum dots, for which spectral instability is a critical barrier for applications in quantum sensing.

## Introduction
Over the last three decades, the range of quantum light sources has greatly expanded. In particular, single-photon or photon-pair emission was demonstrated for a variety of nano-sized emitters including organic molecules, solid-state defects and semiconductor quantum dots (QDs)[1,2]. As such, they form potential building blocks in future quantum-optical technologies, e.g., quantum communication and quantum sensing[3–6]. While the photon statistics itself is the quantum resource in some applications[7–11], most examples rely on quantum interference and thus require coherent radiation and photon indistinguishability[12–14].

To extend the coherence time in the emission of nanoemitters, they are cooled down to cryogenic temperatures to reduce environmental disturbances[15,1,16,17]. Narrow emission linewidths at low temperatures simultaneously offer new opportunities and challenges. For example, sharp emission energies can be applied to sense fluctuations in the micro environment of an emitter[18–20]. The straightforward integration of colloidally synthesized QDs, including the recent emergence of halide-perovskite QDs[21], into biological settings[22,23] and semiconductor devices[3,24,25] offers exciting perspectives for local sensing[26–28].

However, sensing compels coupling to the environment and therefore typically results in sensitivity to unintentional fluctuations in electric and magnetic fields[19,29]. In fact, this sensitivity is often implicated with spectral diffusion (SD), i.e. the temporal variance of the energy of emitted and absorbed photons[30–33]. Such stochastic spectral dynamics currently set considerable limitations on the usability of nanoemitters for quantum applications. Namely, it deteriorates the indistinguishability of emitted photons and hinders the coupling of nanoemitters with cavities, waveguides, and other emitters[34,35].

The quantitative translation of fluctuations in electric field to those of the emission spectrum is described by the quantum-confined Stark effect (QCSE). An external electric bias skews the confining potential within the potential well, resulting in a reduction of both the electron and hole quantization energies (see Fig. 1a). The resulting redshift of the exciton transition follows a parabolic dependence on the electric field magnitude[29,36] (Fig. 1b). Therefore, a rapidly changing electric field results in a momentary variation of the emission and absorption lines (Fig. 1c and Fig. 1d).

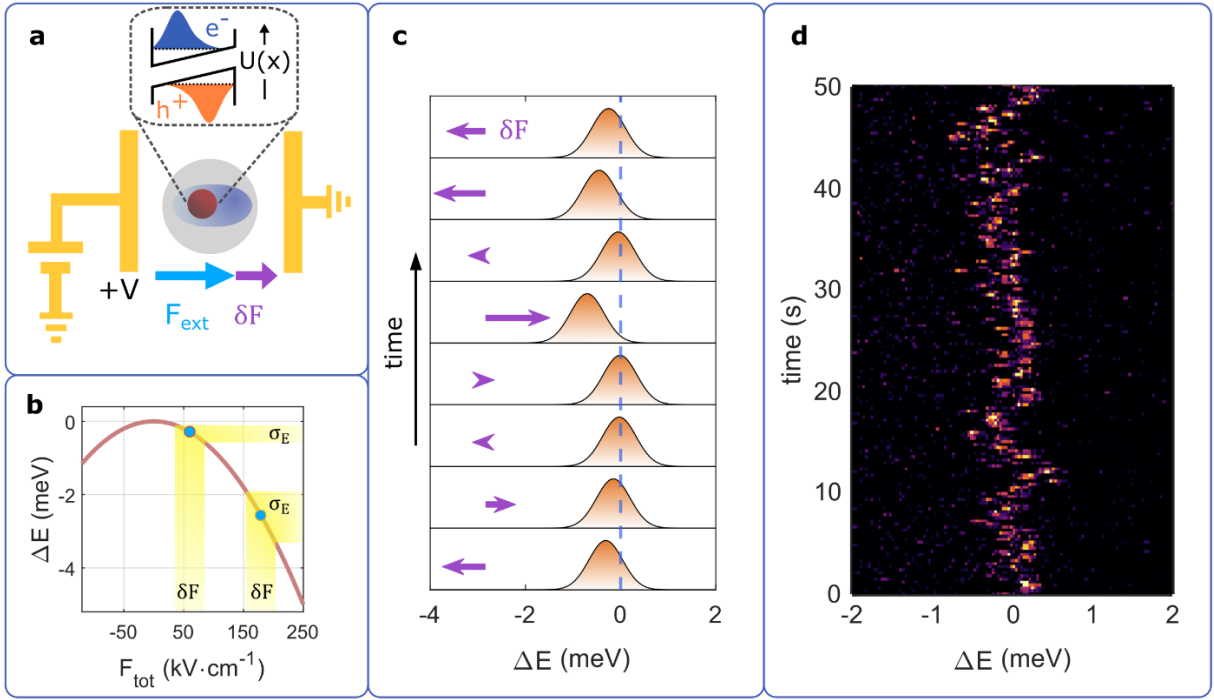

**Fig. 1**: The relation between QCSE and SD. (**a**) Schematic illustration of the sample: single nanorod placed between two electrodes so an external field $F_{ext}$ can be applied in addition to an intrinsic microscopic field $\delta F$. The electric bias causes a distortion of the carrier confinement potential and thereby a reduced transition energy (inset). (**b**) Qualitative example of the redshift due to QCSE. The energy shift depends quadratically on the electric field. For a larger redshift, the same $\delta F$ amplitude induces stronger fluctuations in energy $\sigma_E$ (yellow highlight). (**c**) Schematic illustration of SD caused by QCSE. Random microscopic field fluctuations (purple arrows) cause a redshift that depends quadratically on the field strength. (**d**) Exemplary dataset showing SD. The photoluminescence spectra (each 0.3 seconds integration time) of a single CdSe/CdS core/shell nanorod at T= 6 K exhibit sub-meV irregular fluctuations in energy.

While this causal relation is a prevalent explanation for SD measurements, it is only supported by indirect observations[37]. For example, the PL linewidth in QDs and nitrogen-vacancy centers increases with the electric field magnitude[38,39]. Indeed, a biasing field, increasing the sensitivity of the spectral line to a fluctuating electric field (see Fig. 1b), is a reasonable explanation for these observations. However, elevated temperatures or a strengthened exciton-phonon coupling offer alternative explanations.

The current paper offers the first direct observation of QCSE as the cause of SD in QDs. We show that temporal fluctuations in the PL energy of individual CdSe/CdS dot-in-rod nanoparticles increase the further away they are driven from the QCSE parabola apex. A straightforward quantitative model for QCSE matches our results well. Somewhat surprisingly, we find that an inherent dipole exists in many of the nanoparticles studied here. As a result, a significant improvement in spectral stability can be achieved with a compensating electric field.

## Results

In the experiments below, we study SD by measuring photoluminescence from single CdSe/CdS core/shell nanoparticles in a dot-in-rod geometry. The choice of 3.2 nm core size and 20 nm total nanorod length, on average, sets the energy of the fundamental transition to a wavelength around 590 nm (2.1 eV photon energy)[24]. To enhance chemical and photostability, the inorganic nanocrystals are overcoated with a polystyrene/PMMA nano shell with a thickness of 15 nm. The resulting hybrid organic/inorganic nanoparticles demonstrate superior spectral stability as compared to their uncoated counterparts[40–42].

A custom-built Er:fiber source generates tunable pulses with sub-picosecond duration at a repetition rate of 50 MHz. In order to excite the nano-emitters above the fundamental resonance, the excitation wavelength is set to 540 nm (2.3 eV)[43]. The laser is input into a confocal microscope setup constructed around a bath cryostat (Scientific Magnetics) within which the beam is focused through a 0.9 numerical aperture objective lens (Olympus, M PLAN N, x100 magnification)[44]. A sample containing the hybrid nanoparticles embedded into micro-capacitor structures with a 2 μm gap on a $SiO_2$ substrate (see further details in supplementary notes 1 – 3) is placed in the focal plane of the microscope and held at a temperature of 8 K. The PL collected through the same objective lens is spectrally resolved with a grating monochromator (PI Acton, SP2300, 2400 lines/mm grating) and detected by an electron-multiplying charge-coupled device (EMCCD) camera (Andor, Newton 970). A typical time trace of PL spectra is presented in Fig. 2a with the energy axis centered around the brightest emission line at 2.125 eV. Following previous reports, the three narrow emission lines can be attributed to the F, A1 and A2 transitions in a neutral QD arising from the fine structure splitting of the exciton[45,46]. Within the 90 seconds acquisition time, the three spectral lines fluctuate synchronously within a range of approximately 1.5 meV, indicating that they originate from a single quantum emitter. Integrating over the time axis in Fig. 2a (top inset) all spectral features are significantly blurred, in contrast to the individual 1 s exposure measurements.

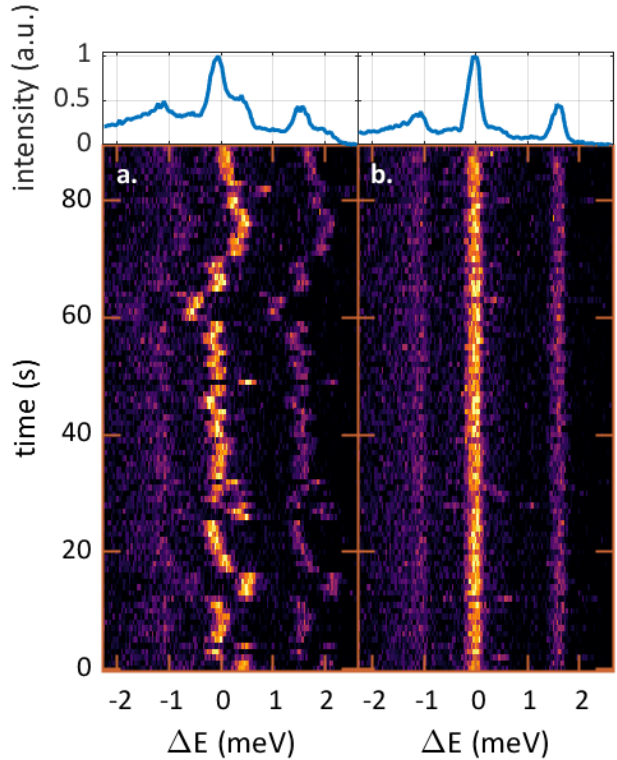

**Fig. 2**: Correcting SD in post processing. (a) 90 consecutive 1s acquisitions of a single QD PL spectrum. Horizontal shifts between the spectra manifest the SD. The top inset presents the integrated spectrum from all measurements. (b) The same dataset corrected in postprocessing with the algorithm. Spectra are now well aligned on the energy axis. The integrated spectrum (top) presents three narrow lines, indicating radiative transitions from the excited states to the ground state.

Alternatively, spectral shifts can be evaluated in post processing in order to realign the spectra with one another (Fig 2b). To this end, a cross-correlation-based algorithm, robust to the naturally low signal-to-noise ratio (SNR) of the spectra, is applied[44]. Subsequently, the three emission lines are clearly aligned in energy. In the integrated spectrum, the linewidth of each feature is similar to that of individual measurements, i.e. less than 0.3 meV. A single-spectra acquisition time of 1 s is selected to obtain reasonable SNR for the correction algorithm while enabling sufficient sampling of the transition-energy dynamics. This procedure provides quantitative information about spectral fluctuations and is used below to statistically analyze SD.

The parabolic dependence of the transition energies on an applied electric field means that both the PL wavelength as well as its sensitivity to field perturbations vary with an external bias. In the regime of small field fluctuations, energy shifts are proportional to the first derivative of the QCSE parabola. As a result, assuming an equal amplitude of field fluctuations ($\delta F$), the range of energy shifts increases with an electric-field induced redshift of the PL (yellow highlight areas in Fig. 1b).

Applying an adjustable DC voltage between -100 V and +100 V, the electric field in the z direction is tuned between -500 kV/cm and +500 kV/cm, respectively. Within this range the PL energy shifts by up to 26 meV, roughly 100 times the spectral linewidth. In contrast to previous reports of nearly complete darkening of the emission[38,47], even under the relatively strong fields used here, the spectrally integrated PL intensity reduces by 40% at most. Due to the dielectric nature of the exciton's environment (CdSe,CdS and polymer encapsulation), the local electric field strength is reduced to approximately 0.35 of the field amplitude in the surrounding vacuum[48] (see supplementary note 4). In the current text and contained graphics, we use only the nominal values of the electric field, i.e. the field amplitude in vacuum.

An initial electric field sweep is performed to determine the PL peak energy as a function of the external electric field (28 voltage steps, 1 s acquisition each). Fig. 3a shows the relative energy shift as a function of the applied electric field ($F_{ext}$) in a range from -400 kV/cm to +350 kV/cm (blue squares). The field dependence of the redshift fits to a parabolic function to a rather high degree (Fig. 3a, orange line)

(1) $$\Delta E = -\frac{1}{2}\beta(F_{ext} - F_{z,0})^2,$$

where $F_{z,0}$ is the projection of the inherent electric bias of the QD on the z axis, defined by the direction of the external field. Such a built-in bias was previously reported for CdSe core-only and CdSe/CdS core/shell QDs both in room-temperature and cryogenic measurements[29,36,38,47,49]. It is assigned to a break of centro symmetry in the exciton state likely due to a combination of asymmetry in the confining potential and strain distribution[36,50]. In the measurement shown in Fig. 3a, the apex offset $F_{z,0}$ is estimated to be 20 kV/cm.

To analyze the interplay between the applied field and spectral dynamics, spectral time traces (100 spectra, 0.5-1 s acquisition time) are taken under several values of the applied field. Fig. 3b presents such a time trace, divided into five sections according to the external bias (dashed white lines). With increasing redshift of the emission line, SD fluctuations clearly intensify in range. For a quantitative evaluation of SD, the peak energy shift in each spectrum is determined via the cross-correlation algorithm. An example of the results, for two temporal sections, is shown in Fig. 3c where the external field is switched from -200 kV/cm to -100 kV/cm at t = 50 s. In each section, the average energy is marked with a blue dashed line and a violet area represents the standard deviation. Indeed, with a red shift of 2 meV the standard deviation in the second section is smaller by a factor of two compared to that of the first section.

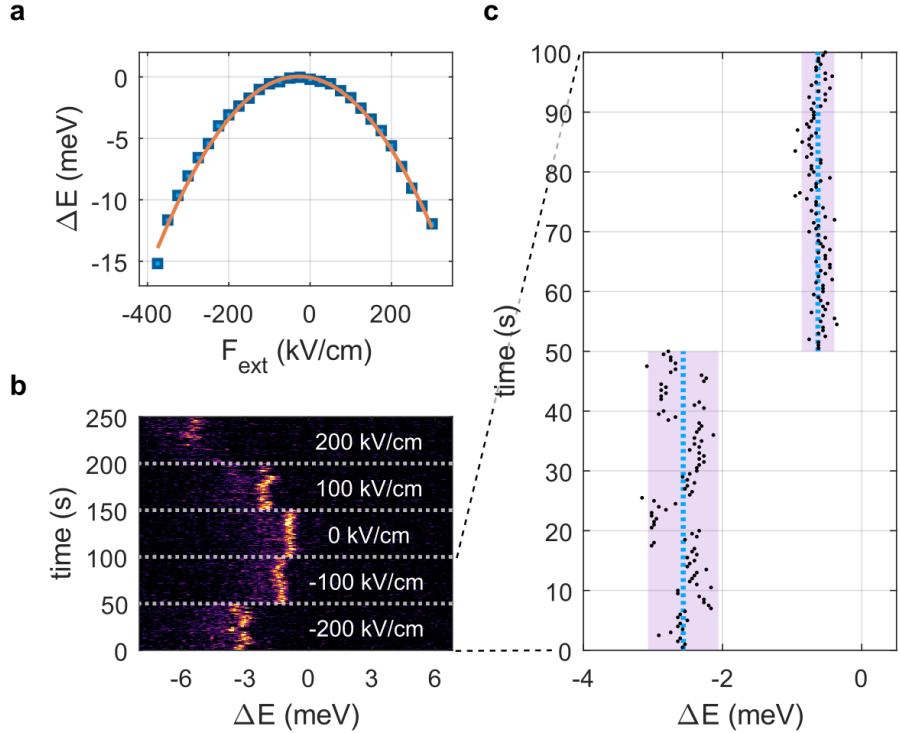

**Fig. 3**: Joint measurement scheme for QCSE and SD. (**a**) The energy shift of the brightest spectral peak vs. nominal field amplitude for a single nanoparticle (blue). A parabolic fit is overlaid in orange. (**b**) Consecutive PL spectra (0.5 s acquisition time) taken for the same QD. The applied external field is changed every 50 s (white dashed lines, field value given alongside). (**c**) Momentary energy shift analyzed for the first 100 s of the measurement in (b). Light blue lines indicate the mean energy shift and violet rectangles indicate the region of two standard deviations to each side of the mean.

Fig. 4 demonstrates this effect for two individual QDs, labeled as QD1 and QD2 from this point onwards. The PL emission energy as a function of the applied electric field, analyzed from a voltage sweep measurement, is shown in Fig. 4a and Fig. 4b for QD1 and QD2, respectively. As expected, the dependence follows a parabolic trend (Eq. (1)) depicted by the orange lines. Notably, the intrinsic offset field parameter, $F_{z,0}$, is positive and large (+520 kV/cm) for QD1 while negative and small (-27 kV/cm) for QD2.

To analyze SD, we conceptually divide the timescale of fluctuations into two regimes: slow dynamics that can be tracked with the 1 s sampling time of our measurement and fast fluctuations which occur within each acquisition. The observable fluctuations in Fig. 3b and Fig. 3c belong to the former whereas the latter manifests as a broadening of the emission line. Fig. 4c and Fig. 4d depict the standard deviation of the emission-peak energy ($\sigma_E$) versus the applied electric field for QD1 and QD2, respectively. For both QDs, the standard deviation, a simple quantitative measure of slow spectral fluctuations, increases with a redshift of the PL, that is, with larger $\Delta F = |F_{ext} - F_{z,0}|$. Importantly, the point of maximal spectral stability is aligned with the apex of the QCSE parabola ($F_{z,0}$), a good indication of the relation between SD and QCSE. In fact, these results qualitatively confirm the picture presented in Fig. 1b – the range of spectral fluctuations increases with further offset from the apex of the parabola. Supplementary Fig. S4 shows similar analyses for two additional nanoparticles that confirm this conclusion.

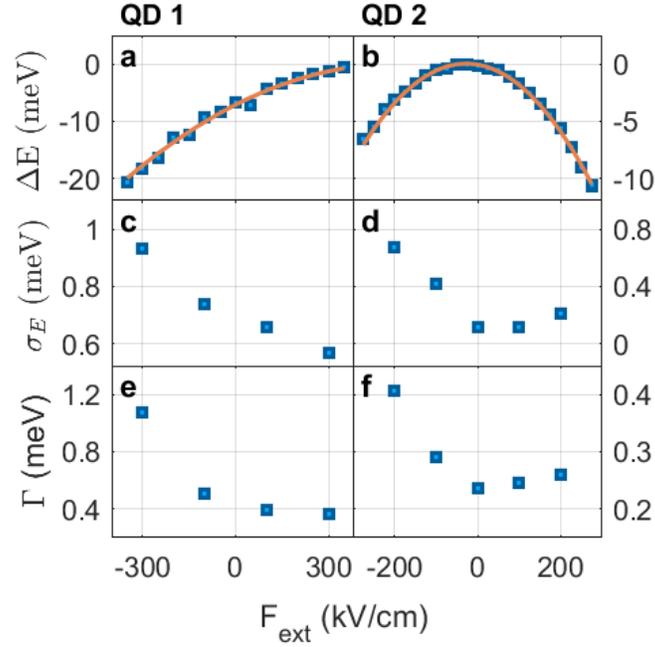

**Fig. 4**: SD measurement for two individual nanoparticles. The energy shift of the PL emission versus the applied electric field for (**a**) QD1 and (**b**) QD2 analyzed from an electric field scan measurement. Orange lines are a parabolic fit for the QCSE following Eq. (1). The standard deviation of PL-emission peak energy analyzed from a time series of spectra, as shown in Fig. 3b, for (**c**) QD1 and (**d**) QD2. Dependence of the spectral width of the emission line, after correction for slow SD, on the external electric field for (**e**) QD1 and (**f**) QD2.

We note that to perform the spectral diffusion analysis as a function of an external electric field we postselect measurements that fulfill several criteria. First, as the spectral-correlation algorithm relies on sufficient SNR, spectra with a particularly low signal level are dismissed. Second, as these measurements require approximately five minutes of laser exposure time, a portion of the measurements were discarded due to photobleaching. Finally, the measured QCSE shift must exhibit a substantial nonlinear dependence so as to vary the sensitivity of the PL energy to field fluctuations. Further details regarding data postprocessing and analysis are given in supplementary note 5. Importantly, none of the measurements featuring a statistically significant result displayed a contradictory trend to that shown in Figs. 3-5 (and Figs. S4 and S6).

In an alternative analysis, focused on fast fluctuations, we align the consecutive spectra within each field-magnitude section and obtain an average spectrum with a high SNR (see supplementary Fig. S5). In order to characterize the emission line shape, we fit each averaged spectrum with a Gaussian function

(2)
$$I(E) = I_0 e^{-\frac{(E-E_0)^2}{\Gamma^2}},$$

where $I(E)$ is the PL spectrum, $E_0$ the energy peak and $\Gamma$ the observed spectral linewidth. Fig. 4e and Fig. 4f present $\Gamma$ as a function of the external-field magnitude for QD1 and QD2, respectively. Similarly to $\sigma_E$, the linewidth decreases when tuning the external field towards the apex of the QCSE parabola ($F_{z,0}$). For a substantial portion of our measurements, the presence of a large inherent bias means that the linewidth can be significantly narrowed by the application of an external field. For QD1, for example, $\Gamma$ at the apex is three times smaller than that without external field.

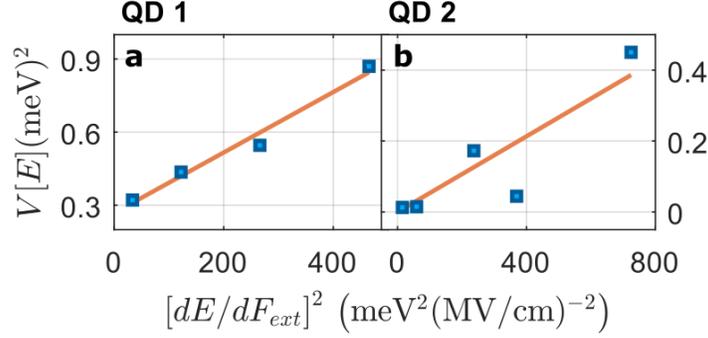

**Fig. 5**: A quantitative analysis of the relation between SD and QCSE. The dependence of slow fluctuations variance ($\sigma_E^2$) on the square of the derivative of the energy with respect to the electric field for (**a**) QD1 and (**b**) QD2. Linear fits (orange lines) indicate that the model presented in Eq. (5) is in a good agreement with our results for parameter values . $\sqrt{\langle \delta F_z^2 \rangle} = 35\ kV/cm$ (QD1) and $\sqrt{\langle \delta F_z^2 \rangle} = 23\ kV/cm$ (QD2).

Indeed, previous studies already pointed out this relation for both QDs and nitrogen-vacancy centers in diamond, supporting a connection between QCSE and SD[38,39]. However, the data shown in Fig. 4c and Fig. 4d provides a more direct observation - a correlation between spectral dynamics and the applied field. That is, we observe the energetic dynamics itself rather than its consequence on a static measurement. As mentioned above, a few alternative explanations can account for the broadening of PL emission lines in a steady state, such as thermal effects and enhanced phonon coupling. These cannot provide an explanation for the extended range of slow spectral fluctuations shown in this work.

To further support the mechanistic explanation for SD described above, we compare the results of Fig. 4c,d to a straightforward statistical model. Including a local fluctuating field $\delta \vec{F}(t)$ in Eq. (1), the PL transition energy as a function of time is expressed as

$$(3) \quad E(t) = E_0 - \frac{1}{2}\beta\left(F_{ext}\hat{z} + \delta \vec{F}(t) - \vec{F}_0\right)^2.$$

The field fluctuations can be divided into parallel and orthogonal axes $\delta \vec{F} = \delta \vec{F}_\perp + \delta F_z \cdot \hat{z}$ with respect to the applied field. Assuming that fluctuations in orthogonal axes are independent of one another,

$$(4) \quad V[E] = \sigma_E^2 \equiv \langle (E(t) - \langle E \rangle)^2 \rangle = V_0 + \langle \delta F_z^2 \rangle \cdot \beta^2 (F_{ext} - F_{z,0})^2,$$

where $V_0$ consists of terms that contribute to the variance but do not depend on the strength of the external electric field, $F_{ext}$ (see supplementary note 6 for the full derivation). In order to fit this model to our data we note that $\left[dE/dF_{ext}\right]^2 = \beta^2(F_{ext} - F_{z,0})^2$ which allows us to rewrite equation (4) as

$$(5) \quad V[E] = V_0 + \langle \delta F_z^2 \rangle \cdot \left[dE/dF_{ext}\right]^2.$$

Intuitively, equation (5) merely reflects the fact that spectral fluctuations are proportional to the absolute value of the derivative ($dE/dF_{ext}$) and therefore the variance increases with its second power.

In Fig. 5a and Fig. 5b the same data as in Fig. 4c and Fig. 4d is plotted against $[dE/dF_{ext}]^2$, respectively. Two additional datasets are presented in supplementary Fig. S6. Orange lines portray the linear fits of the data according to equation (5). From these we obtain a unique estimate for the standard deviation of the local electric field averaged over the four measurements presented in the paper and supplementary information - 9±2.5 kV/cm. We note, that this value reflects field amplitude within the CdSe core of the nanoparticle – multiplied by a factor of 0.35 due to the dielectric environment (see supplementary note 5). The very similar values for $\delta F_z$ for all analyzed nanoparticles strengthens the validity of the simplified general model applied here.

## Summary

To summarize, the current work provides the first direct observation that photon energy fluctuations in the PL of individual colloidal QDs result from spurious electric fields in their micro environment. Applying an external field, we find that the photon energy variance increases with the red shift of the emission line. A quantitative fit of this dependence to a straightforward model estimates the standard deviation of the electric field sensed by the exciton as 9 kV/cm on average.

A continuous improvement in spectral stability is a crucial step towards the application of semiconductor nanocrystals as local quantum sensors of their environment. A direct consequence of this study is that the spectral stability of single QDs, often biased by an inherent field, can be substantially improved by applying an inverse unbiasing field. To further improve spectral stability, the method presented here can provide spectroscopic feedback for the optimization of synthesis and sample-preparation procedures for quantum applications. Performing such studies with a higher temporal and spatial resolution, taking advantage of new detector technology[10] and nanocrystal architectures[51,52], could play an important role in isolating the microscopic source of electric field disturbances.

## Associated content

**Supporting information**

The supporting text and figures include further details about nanocrystal synthesis, sample preparation, data processing and mathematical modeling.

## Author information

**Author contribution**

F. Conradt and V. Bezold contributed equally to this work.

**Notes**

The authors declare no competing financial interest.

## Acknowledgement

The authors acknowledge funding by the Deutsche Forschungsgemeinschaft (DFG, German Research Foundation) – Project-ID 425217212 – SFB 1432. R. T. thanks the Minerva foundation for their support. The authors express gratitude to Matthias Hagner for assisting with the nanofabrication processes.

Supporting information for "Electric-field fluctuations as the cause of spectral instabilities in colloidal quantum dots"

Frieder Conradt[1,2], Vincent Bezold[1,2], Volker Wiechert[1], Steffen Huber[3], Stefan Mecking[3], Alfred Leitenstorfer[1], Ron Tenne[1]

[1] Department of Physics and Center for Applied Photonics, University of Konstanz, D-78457 Konstanz, Germany

[2] These authors contributed equally

[3] Chair of Chemical Materials Science, Department of Chemistry, University of Konstanz, D-78457 Konstanz, Germany

## Supplementary note 1: Capacitor Preparation

To apply an external electric field on single nanocrystals, we prepare an interdigitated gold-electrode structure with a gap size of 2 µm. The electrode structure is fabricated via electron-beam lithography (EBL) on a crystalline SiO2 substrate. To do so, the substrate is coated with a 220 nm layer of PMMA (A4) resist and a 5 nm aluminum layer to prior to the patterning. Developer fluid is a mixture of isopropanol, MIBK methyl isobutyl ketone (MIBK) and methyl ethyl ketone (MEK) (ratio: 3:1:0.06) in which the exposed sample is immersed for 60 seconds. Subsequently, a layer of 5 nm Chromium (Cr) for enhanced adhesion and 60 nm of gold are deposited on the developed sample by thermal evaporation. By placing the sample for 12 hours or more in an acetone bath, the remaining resist and the corresponding metal layer is lifted off. The sample with the remaining electrodes is mounted on a circuit board and connected with bond wires. The gap size was chosen to enable the application of a strong field magnitude of up to 500 kV/cm using a reasonable analog voltage of 100 V. An optical microscope image of the final structure is shown in Fig. S1.

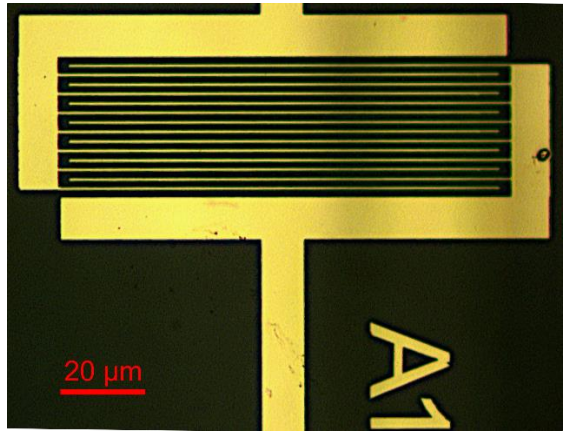

**Fig. S1**. An optical-microscope image (edited) of a gold electrode structure on a SiO2 substrate. The structure is fabricated with electron-beam lithography. The gap between adjacent metallic strips size is about 2 µm.

The precise gap size is measured with an atomic force microscope (AFM). Such a measurement is shown in Fig. S2. The average gap size if 1.85(5) µm.

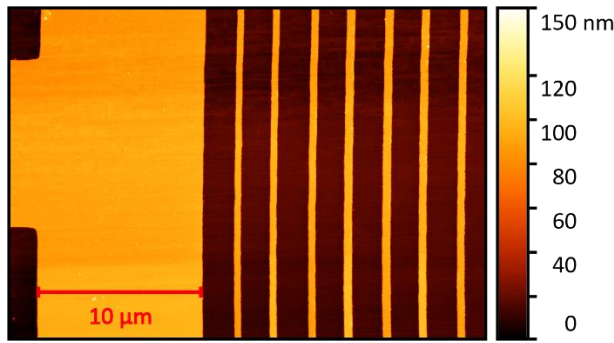

**Fig. S2**. Atomic-force-microscope (AFM) image of a capacitor structure (same as in Fig. S1). The average gap size is of 1.85(5) µm.

## Supplementary note 2: Quantum dots deposition

Embedding quantum dots (QDs) into the capacitor structure is done by a drop-casting process. Prior to this, the sample is placed in a plasma cleaner to reduce droplet formation during drop casting. A water based dispersion of hybrid nanoparticles (see Supplementary note 3) is diluted into ethanol in a ratio of 1:50. A 5 µl drop of this solution is then deposited on the sample. For some samples, this step was repeated to increase the concentration of nanoemitters. The ethanol solvent supports the formation of a thin liquid film rather than water droplets and leads to a faster evaporation which guarantees an even distribution of the nanoparticles. The distribution of functional nanoparticles is verified with a photoluminescence (PL) imaging microscope. Fig. S3 shows a capacitor structure with embedded nanoemitters. Backlight illumination enables the observation of both the structure (dark portions) and the photoluminescence of the QDs. In Fig. S3, the digital saturation of the PL signal is done intentionally to highlight the dimmer scattered light from the substrate surface.

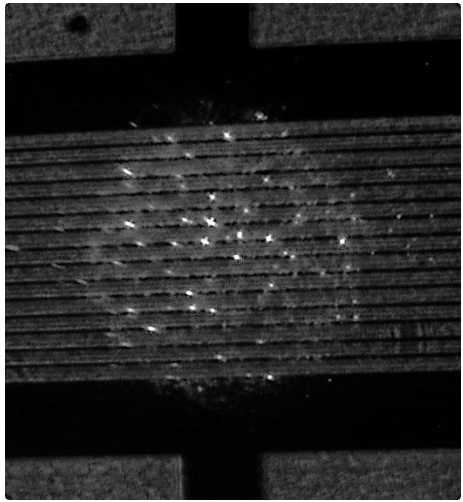

**Fig. S3.** PL of nanoemitters embedded in a capacitor structure. Due to backlight illumination, the metallic structure appears dark while scattered PL light from the substrate surface exposes nonmetallic portions. The image not only shows that QDs are embedded within an electrode gap but also helps to ensure that the emitters do not cluster and can be observed individually.

## Supplementary Note 3: Synthesis of encapsulated nanorods

Full details regarding the synthesis of nanorods (NRs) used the reader is referred to reference 1. In this section, we provide a summary of this procedure.

The following materials were used in the synthesis: Cadmium oxide (> 99.99%-Cd, lot MKBT7524V), hexylphosphonic acid ( 95%, lot MKBX1133V), oleylamine (70%), oleic acid (90%), octanethiol (> 98.5%), tri-*n*-octylphosphine (TOP; 99%), tri-*n*-octylphosphine oxide (TOPO; 99%), methyl methacrylate (MMA) (99%, contains < 30 ppm MEHQ as inhibitor), *p*-divinylbenzene (85%, stabilized with 4-*tert*-butylpyrocatechol), styrene (> 99%, contains 4-tert-butylcatechol as stabilizer), 2,2'-azobis(2-methylpropionitrile) (98%) and sodium dodecyl sulfate (> 99%, dust free pellets) and octadecene (90%) were purchased from Sigma-Aldrich. *n*-Octadecylphosphonic (ODPA, 98%, lot 807601N16) acid was obtained from PCI.

Standard organic solvents and chemicals were obtained from various commercial suppliers such as Sigma-Aldrich, ABCR, VWR and Roth.

Deionized water was distilled under a nitrogen atmosphere. Methyl methacrylate was filtered over basic aluminum oxide, dried over 4 Å molecular sieves, degassed by three freeze-pump-thaw cycles and stored inside a glovebox at -30 °C. Styrene was vacuum transferred, degassed by three freeze-pump-thaw cycles and stored inside a glovebox at -30 °C.

### Synthesis of CdSe cores

The synthesis of the CdSe core particles follows a modified protocol by Carbone *et al.*[2]. Shortly, 60 mg of CdO, 280 mg of octadecylphosphonic acid and 3000 mg of tri-n-octylphosphine oxide (TOPO) were mixed in a three-neck flask. Under vacuum and at a temperature of 150°C the reagents were degassed for one hour. Under a nitrogen environment, the temperature was increased to 330°C and the solution was stirred for around two hours. The mixture was stirred until it turned clear, indicating the complexation of the $Cd^{2+}$ ions. Subsequently 1.8 mL of tri-n-octylphosphine (TOP) was added and the mixture heated up to 370°C. A TOP-Se solution is prepared by 57 mg of Se and 360 mg of tri-n-octylphosphine by stirring for one hour at room temperature. This solution is then injected, and the reaction cooled down as soon as the desired size of the CdSe seeds is reached. The typical growth time is about 30 to 60 seconds. An amount of 3 mL of toluene was added at a temperature of 100 °C. Afterwards the cores were purified by precipitation in methanol, centrifugation, discarding the supernatant and redispersion in toluene. This is repeated three times. Finally, the cores were separated, dried and redispersed in 4 mL of *n*-hexane.

### Synthesis of CdSe/CdS/CdS rods

CdSe/CdS seeded nanorods were prepared following the synthesis procedure published by Carbone *et al.* and upscaled by a factor of 1.5[2]. In a following step, a second overcoat shell of CdS was added, realizing CdSe/CdS/CdS core/shell/shell. The recipe for this synthesis was adapted from the one described by Coropceanu *et al.*[3] which uses the seeded CdSe/CdS nanorods as its starting point.

20 nmol of the CdSe/CdS rods in hexane are mixed with 1.5 mL of octadecene, 1.5 mL of oleylamine and 1.5 mL of oleic acid. Residues of water and hexane were removed for 45 minutes at a temperature of 50°C and 15 minutes at 105°C in vacuum. The final reaction is realized at a temperature 310°C and under a nitrogen atmosphere. The injection of the precursor solutions (3 mL of Cd-oleate in octadecene and 3 mL of 1-octanethiol in octadecene) already started when reaching 210°C. Both solutions were added at a rate of 1.5 mL/h (theoretically 1 monolayer of CdS per hour). The desired shell thickness (here: 2 mono layers) and the size of the initial CdSe/CdS NRs define the amount of precursor solution. Assuming a complete conversion of the Cd-oleate, the ratio of octanethiol to Cd was chosen to be 1.2:1. After the precursor was added, the reaction was cooled down and precipitated in acetone/MeOH (70:30). The NRs were extracted via centrifugation, dried, and redispersed in toluene.$\sigma$

Polymer encapsulation of nanorods

In the next synthetic step, the NRs were embedded into cross-linked polystyrene spheres overcoated with an additional cross-linked polymethyl methacrylate (PMMA) shell. The polystyrene encapsulation is based on a procedure by De San Luis *et al.*[4] . The aqueous phase is prepared with 42 mg of sodium dodecyl sulfate (SDS) and 21 mg of $NaHCO_3$, both degassed and dissolved in 25 mL of distilled water. The organic phase consists of 2.3 mL of styrene, 84 µL of *n*-hexadecane, 21 µL of 1,4-divinylbenzene and 0.6 mL (5 nmol) of the NRs dispersed in toluene. After stirring, the organic phase is added to the aqueous phase. A miniemulsion was prepared by ultrasonication for 4 minutes at 80% intensity, while the mixture was cooled in an ice batch to prevent polymerization. A mixture of 21 mg of SDS and 10 mL of distilled water was added afterwards to the emulsion, which was subsequently heated up to 75 °C. Polymerization was initiated by adding 11 mg of potassium peroxydisulfate, dissolved in 4 mL of distilled water. The polymerization process lasted for 6 hours.

In a final step, an additional cross-linked PMMA shell (theoretical thickness of 15 nm) was added to coat the nanoparticles. For this purpose, 2 mg of AIBN were added to 20 mL of the polystyrene dispersion After heating up the mixture to 75°C , methyl methacrylate (MMA) mixed with 1,4-divinylbenzene (100:1 weight ratio) was added via a syringe pump with an injection rate of 0.25 mL/min. The mixture was stirred for 3 hours. The required quantity of MMA was calculated by the solids content of the dispersion and size (DLS number average) of the initial particles and the desired shell thickness assuming the polystyrene particles to have a density of 1.05 g/mL.

## Supplementary note 4: Estimation of the effective electric field

Due to the permittivity of the polystyrene and CdS shells, the effectively electric field sensed by the exciton, confined to the core, is a fraction of the applied field strength $E_a = \frac{U_a}{d}$ in vacuum with voltage bias $U_a$ and gap size $d$. The effective electric field in a dielectric sphere placed in an external field can be estimated as[6]

$$E_{eff} = \frac{3 \cdot \varepsilon_{gap}}{\varepsilon_{sphere} + 2 \cdot \varepsilon_{gap}}. \quad \text{S1}$$

We obtained a value of $E_{QD} = 0.35 \cdot E_{applied}$, using $\varepsilon_{PS} = 2.5$ and $\varepsilon_{CdS} \approx 9.2$[9] as the value of permittivity for polystyrene and CdS respectively. Here, the effect of the small CdSe core is neglected as its permittivity ($\varepsilon_{CdSe} \approx 8.5$) nearly matches that of the CdS shell.

Supplementary Fig. S4: Spectral diffusion under an external electric field for two additional QDs

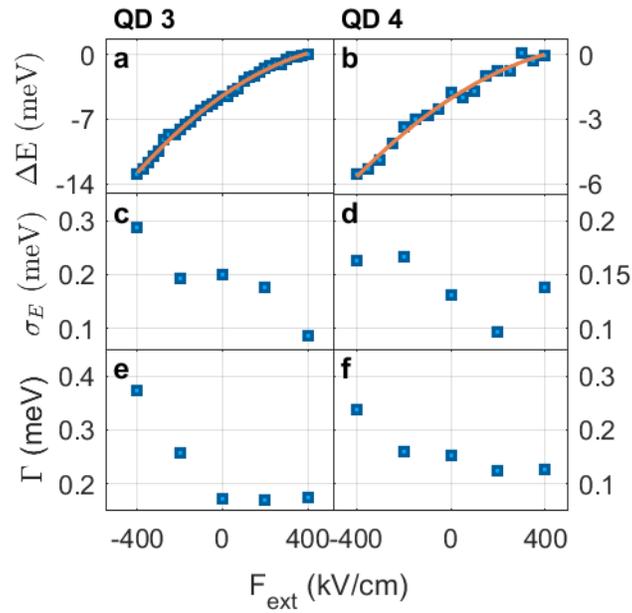

**Fig. S4**. Spectral diffusion versus electric field for two additional individual QDs beyond the data presented in the main text. This figure is identical in format to Fig. 4 of the main text. The energy shift of the PL emission versus the applied electric field for (a) QD3 and (b) QD4 analyzed from an electric field scan measurement. Orange lines are a parabolic fit for the QCSE. The standard deviation of PL-emission peak energy analyzed from a time series of spectra, as shown in Fig. 3b, for (c) QD3 and (d) QD4. Dependence of the spectral width of the emission line, after correction for slow SD, on the external electric field for (e) QD3 and (f) QD4.

## Supplementary Note 5: Preprocessing of measurement data

The principal challenge to quantify possibly fast spectral fluctuations is to precisely measure shifts of spectra despite a relatively low per-pixel signal-to-noise ratio (SNR). The method used here is based on a cross-correlation algorithm of $N_{spec}$ consecutive spectra with ~1 s acquisition time each[5] and is similar to algorithms used to obtain high quality images in astrophotography (lucky imaging). Even though PL emission at cryogenic temperatures is significantly more stable in comparison with room temperature luminescence, the presence of intensity fluctuations still reduces the performance of the correlation algorithm. Therefore, in an initial step, we filter out spectra in which the total intensity is more than three standard deviations below the mean over all spectra.

Additional preprocessing steps mitigate the effect of spectral instabilities that are not attributed to spectral diffusion, such as cosmic rays and charging. Single spectra whose peaks shift by more than three standard deviations from the mean peak position are excluded from further analysis. We note, however, that such events are hardly present in the data used in the paper.

On rare occasions, irreversible sudden spectral jumps are observed. Since we suspect that such sudden shifts are due to, e.g., changes in the chemical environment of the nano particle rather than SD, we disregard these in the analysis of spectral shifts. To identify sudden and large spectral jumps, the vector of spectral positions is convoluted with a step function:

$$\vec{S} = \vec{a} * \Delta\vec{E}, \qquad \text{S2}$$

where $\vec{a}$ is defined as $\vec{a} = (-1, -1, \ldots, -1, 1, 1, \ldots, 1)$ with the length 2*n (where n is typically 2-3) and the symbol $*$ stands of the convolution operator. Spectral jumps in $\Delta\vec{E}$ result large absolute values in $\vec{S}$. Thereby, the dataset can be divided in several segments between such events. In a following step, the mean and variance are estimated within each segment separately. The variance is then averaged between all segments to obtain a quantitative estimate of spectral fluctuations under a constant external electric field.

Supplementary Fig. S6: Analysis of emission linewidth

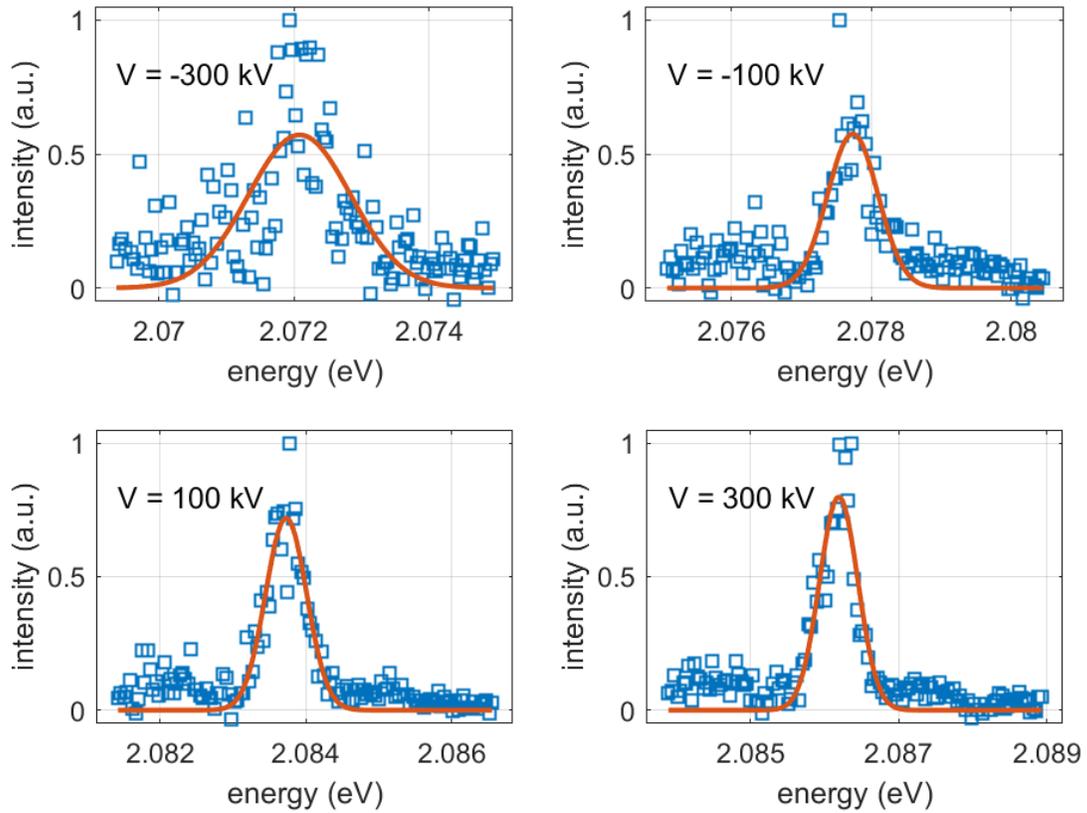

**Fig. S5**. An example demonstrating the analysis of spectral linewidth for QD1. Panels present the normalized PL spectra for QD1 under four different voltage values (indicated on axes). Each spectrum is fitted with a single Gaussian function (red line). The values for the width parameter $\Gamma$, presented in Fig. 4 of the main text, are taken from these fits. We note that outlying high intensities within some spectra are ignored in the fits as they do not belong to the continuous spectral lineshape. While uncertain, we tend to attribute these to systematic noise in the camera rather than to realistic spectral features.

## Supplementary note 6: Deriving the dependence of SD on an applied electric field

This section details the numerical model used to fit the dependence of the variance of the PL energy on the externally applied electric field, presented in Fig. 5 of the main text. The relation between the two is formed through the QCSE and the presence of electric-field fluctuation in the micro environment of the QD.

We begin by considering the transition energy from the ground state to the exciton state ($E$) and its dependence on a general electric field ($\vec{F}$). According to QCSE

$$E = E_0 - \frac{1}{2}\beta(\vec{F} - \vec{F}_0)^2, \quad \text{S3}$$

where $E_0$ is the transition energy in the absence of any electric field and $\beta$ is the polarizability – a property of the material and the structure of the QD. The model includes a built-in electric-field vector $\vec{F}_0$ that is time independent. The presence of such a field component had been suggested by repeated observation of an offset in the energy maximum in QCSE experiments on multiple type of emitters[7,8]. The environmental electric-field vector ($\vec{F}$) can be further divided into two parts: an externally applied field ($\vec{F}_{ext}$) and a fluctuating component that results from the microscopic environment of the QD ($\delta\vec{F}(t)$). Without loss of generality, we set the external electric field in the $\hat{z}$ direction so that $\vec{F}_{ext} = F_{ext}\hat{z}$, yielding

$$E(t) = E_0 - \frac{1}{2}\beta(\delta F_\parallel + F_{ext} - F_{0,\parallel})^2 - \frac{1}{2}\beta(\delta\vec{F}_\perp - \vec{F}_{0,\perp})^2, \quad \text{S4}$$

where we have separated the fluctuating field into two components - parallel and orthogonal with respect to the external field.

By definition of the statistics for the fluctuating field

$$\langle \delta F_\parallel \rangle = 0 \,;\; \langle \delta\vec{F}_\perp \rangle = 0. \quad \text{S5}$$

Taking the time average of Eq. S4 under the assumption of ergodicity and plugging in Eqs. S5 we obtain

$$\langle E \rangle = E_0 - \frac{\beta}{2}\left[\Delta F_\parallel^2 + \langle \delta F_\parallel^2 \rangle + \langle \delta\vec{F}_\perp^2 \rangle\right]. \quad \text{S6}$$

Our analysis targets the effect of the fluctuating field on the variance of the transition energy

$$V[E] \equiv \langle (E - \langle E \rangle)^2 \rangle = V_0 + \beta^2 \langle \delta F_\parallel^2 \rangle \cdot (F_{ext} - F_{0,\parallel})^2, \quad \text{S7}$$

where $V_0$ contains multiple terms that do not depend on $F_{ext}$ - the parameter varied in our experiments. In the last step, we have assumed that

$$\langle \delta F_\parallel^3 \rangle = 0, \quad \text{S8}$$

enforcing reversal symmetry on the fluctuations and

$$\langle \delta F_\parallel \delta\vec{F}_\perp^2 \rangle = 0, \quad \text{S9}$$

relying on a lack of correlation between the electric-field fluctuations in orthogonal directions. A violation of one of these assumption leads to a term with a linear dependence on $F_{ext} - F_{0,\parallel}$ in Eq. S7.

To simplify the comparison of this model to the experimental results, we note that

$$\frac{\partial \langle E \rangle}{\partial F_{ext}} = \beta(F_{ext} - F_{0,\parallel}). \quad \text{S10}$$

Using the last expression in Eq. S7, we achieve the main result of this section,

$$V[E] = V_0 + \left(\frac{\partial \langle E \rangle}{\partial F_{ext}}\right)^2 \cdot \langle \delta F_\parallel^2 \rangle. \qquad \text{S11}$$

Fig. 5 of the main text and Fig. S5 of the supplementary information present the comparison of the expression in Eq. S*11* to the experimental data of 4 different individual quantum dots. The fit parameters here are $V_0$ and $\langle \delta F_\parallel^2 \rangle$. The first contains multiple contributions due to fluctuations in the field component that is orthogonal to the electric field. In addition, the noise in the measurement of the PL spectrum also contributes to external-field-independent fluctuations in the PL energy peak, effectively added to $V_0$.

## Supplementary Fig. S6: Numerical modeling of SD for two additional QDs

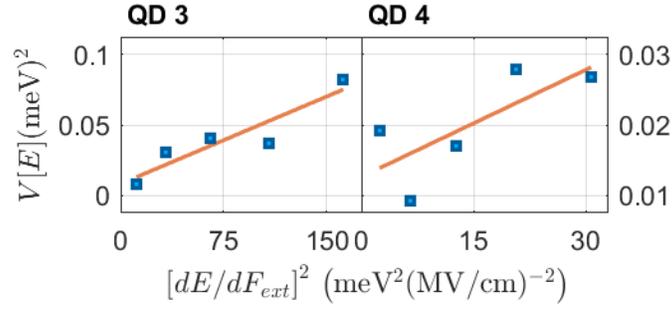

Fig. S6: A quantitative analysis of the relation between SD and QCSE for two additional single-QD measurements. This figure is identical in format to Fig. 5 of the main text. The dependence of slow fluctuations variance ($\sigma_E^2$) on the square of the derivative of the energy with respect to the electric field for QD3 (left) and QD4 (right). Linear fits (orange lines) indicate that the simplistic model presented in Eq. (5) of the main text is in a good agreement with our results. $\sqrt{\langle \delta F_z^2 \rangle} = 20 \; kV/cm$ (QD3) and $\sqrt{\langle \delta F_z^2 \rangle} = 23 \; kV/cm$ (QD4).